\input harvmac
%
%
\font\ticp=cmcsc10
%
%
\lref\Page{
  D.~N.~Page,
  ``Information in black hole radiation,''
  Phys.\ Rev.\ Lett.\  {\bf 71}, 3743 (1993)
  [arXiv:hep-th/9306083].
}
\lref\tHooholo{
  G.~'t Hooft,
  ``Dimensional reduction in quantum gravity,''
  arXiv:gr-qc/9310026.
  }
\lref\Sussholo{
  L.~Susskind,
  ``The World as a hologram,''
  J.\ Math.\ Phys.\  {\bf 36}, 6377 (1995)
  [arXiv:hep-th/9409089].
}
\lref\BoussoJU{
  R.~Bousso,
  ``The holographic principle,''
  Rev.\ Mod.\ Phys.\  {\bf 74}, 825 (2002)
  [arXiv:hep-th/0203101].
}
\lref\Mald{
  J.~M.~Maldacena,
  ``The large N limit of superconformal field theories and supergravity,''
  Adv.\ Theor.\ Math.\ Phys.\  {\bf 2}, 231 (1998)
  [Int.\ J.\ Theor.\ Phys.\  {\bf 38}, 1113 (1999)]
  [arXiv:hep-th/9711200].
}
\lref\GiddingsQU{
  S.~B.~Giddings,
  ``The boundary S-matrix and the AdS to CFT dictionary,''
  Phys.\ Rev.\ Lett.\  {\bf 83}, 2707 (1999)
  [arXiv:hep-th/9903048].
}
\lref\HorowitzVP{
  G.~T.~Horowitz,
 ``Tachyon condensation and black strings,''
  JHEP {\bf 0508}, 091 (2005)
  [arXiv:hep-th/0506166].
}
\lref\HorowitzMR{
  G.~T.~Horowitz and E.~Silverstein,
  ``The inside story: Quasilocal tachyons and black holes,''
  Phys.\ Rev.\  D {\bf 73}, 064016 (2006)
  [arXiv:hep-th/0601032].
}
\lref\BHMR{
  S.~B.~Giddings,
  ``Black holes and massive remnants,''
  Phys.\ Rev.\ D {\bf 46}, 1347 (1992)
  [arXiv:hep-th/9203059].
}
\lref\EHNS{
  J.~R.~Ellis, J.~S.~Hagelin, D.~V.~Nanopoulos and M.~Srednicki,
  ``Search For Violations Of Quantum Mechanics,''
  Nucl.\ Phys.\  B {\bf 241}, 381 (1984).
}
\lref\Hawkunc{
  S.~W.~Hawking,
  ``Breakdown Of Predictability In Gravitational Collapse,''
  Phys.\ Rev.\ D {\bf 14}, 2460 (1976).
}
\lref\EinsDov{A. Einstein, {\sl Relativity: the special and general theory} (Dover, New York).}
\lref\BryceI{
B. DeWitt, ``The Quantization of Geometry'', in
{\it Gravitation: An Introduction to Current Research},  ed. Witten L (New
York, Wiley, 1962).}
\lref\BryceII{
  B.~S.~DeWitt,
  ``Quantum Theory Of Gravity. 1. The Canonical Theory,''
  Phys.\ Rev.\  {\bf 160}, 1113 (1967).
}
\lref\SGIII{
  S.~B.~Giddings,
  ``(Non)perturbative gravity, nonlocality, and nice slices,''
  Phys.\ Rev.\  D {\bf 74}, 106009 (2006)
  [arXiv:hep-th/0606146].
}
\lref\PageUC{
  D.~N.~Page and W.~K.~Wootters,
 ``Evolution Without Evolution: Dynamics Described By Stationary
  Observables,''
  Phys.\ Rev.\  D {\bf 27}, 2885 (1983).
}
\lref\BanksCW{
  T.~Banks,
 ``T C P, Quantum Gravity, The Cosmological Constant And All That..,''
  Nucl.\ Phys.\  B {\bf 249}, 332 (1985).
}
\lref\HartleGN{
  J.~B.~Hartle,
  ``Prediction in quantum cosmology,''
{\it  In *Cargese 1986, Proceedings, Gravitation in Astrophysics*, 329-360. }
}
\lref\RovelliPI{
  C.~Rovelli,
  ``Quantum reference systems,''
  Class.\ Quant.\ Grav.\  {\bf 8}, 317 (1991).
}
\lref\RovelliPH{
  C.~Rovelli,
  ``What is observable in classical and quantum gravity?,''
  Class.\ Quant.\ Grav.\  {\bf 8}, 297 (1991).
}
\lref\RovelliJM{
  C.~Rovelli,
  ``Quantum mechanics without time: a model,''
  Phys.\ Rev.\  D {\bf 42}, 2638 (1990).
}
\lref\RovelliBZ{
  C.~Rovelli,
  ``Partial observables,''
  Phys.\ Rev.\  D {\bf 65}, 124013 (2002)
  [arXiv:gr-qc/ 0110035].
}
\lref\TsamisYU{
  N.~C.~Tsamis and R.~P.~Woodard,
  ``Physical Green's functions in quantum gravity,''
  Annals Phys.\  {\bf 215}, 96 (1992).
}
\lref\SmolinIS{
  L.~Smolin,
  ``Finite diffeomorphism invariant observables in quantum gravity,''
  Phys.\ Rev.\  D {\bf 49}, 4028 (1994)
  [arXiv:gr-qc/9302011].
}
\lref\AshtekarWB{
  A.~Ashtekar, R.~Tate and C.~Uggla,
  ``Minisuperspaces: Observables and quantization,''
  Int.\ J.\ Mod.\ Phys.\  D {\bf 2}, 15 (1993)
  [arXiv:gr-qc/9302027].
}
\lref\MarolfWH{
  D.~Marolf,
  ``Quantum observables and recollapsing dynamics,''
  Class.\ Quant.\ Grav.\  {\bf 12}, 1199 (1995)
  [arXiv:gr-qc/9404053].
}
\lref\MarolfNZ{
  D.~Marolf,
 ``Almost ideal clocks in quantum cosmology: A Brief derivation of time,''
  Class.\ Quant.\ Grav.\  {\bf 12}, 2469 (1995)
  [arXiv:gr-qc/9412016].
}
\lref\AmbjornMK{
  J.~Ambjorn, K.~N.~Anagnostopoulos, U.~Magnea and G.~Thorleifsson,
 ``Geometrical interpretation of the KPZ exponents,''
  Phys.\ Lett.\  B {\bf 388}, 713 (1996)
  [arXiv:hep-lat/9606012].
}
\lref\AmbjornWC{
  J.~Ambjorn and K.~N.~Anagnostopoulos,
  ``Quantum geometry of 2D gravity coupled to unitary matter,''
  Nucl.\ Phys.\  B {\bf 497}, 445 (1997)
  [arXiv:hep-lat/9701006].
}
\lref\GambiniYJ{
  R.~Gambini, R.~Porto and J.~Pullin,
  ``Fundamental decoherence from quantum gravity: A pedagogical review,''
  arXiv:gr-qc/0603090.
}
\lref\GambiniPE{
  R.~Gambini, R.~Porto and J.~Pullin,
  ``A relational solution to the problem of time in quantum mechanics and
  quantum gravity induces a fundamental mechanism for quantum  decoherence,''
  New J.\ Phys.\  {\bf 6}, 45 (2004)
  [arXiv:gr-qc/0402118].
}
\lref\DittrichCB{
  B.~Dittrich,
 ``Partial and complete observables for Hamiltonian constrained systems,''
  arXiv:gr-qc/0411013.
}
\lref\DittrichKC{
  B.~Dittrich,
  ``Partial and Complete Observables for Canonical General Relativity,''
  Class.\ Quant.\ Grav.\  {\bf 23}, 6155 (2006)
  [arXiv:gr-qc/0507106].
}
\lref\ThiemannWK{
  T.~Thiemann,
  ``Reduced phase space quantization and Dirac observables,''
  Class.\ Quant.\ Grav.\  {\bf 23}, 1163 (2006)
  [arXiv:gr-qc/0411031].
}
\lref\PonsRZ{
  J.~M.~Pons and D.~C.~Salisbury,
  ``The issue of time in generally covariant theories and the Komar-Bergmann
  approach to observables in general relativity,''
  Phys.\ Rev.\  D {\bf 71}, 124012 (2005)
  [arXiv:gr-qc/0503013].
}
\lref\GMH{
  S.~B.~Giddings, D.~Marolf and J.~B.~Hartle,
  ``Observables in effective gravity,''
  arXiv:hep-th/0512200.
}
\lref\GaGi{
  M.~Gary and S.~B.~Giddings,
  ``Relational observables in 2d quantum gravity,''
  arXiv:hep-th/0612191.
}
\lref\SGDM{
  S.~B.~Giddings and D.~Marolf,
  ``A global picture of quantum de Sitter space,''
  arXiv:0705.1178 [hep-th],
and work in progress.}
\lref\BPS{
  T.~Banks, L.~Susskind and M.~E.~Peskin,
  ``Difficulties For The Evolution Of Pure States Into Mixed States,''
  Nucl.\ Phys.\ B {\bf 244}, 125 (1984).
}
\lref\AmatiWQ{
  D.~Amati, M.~Ciafaloni and G.~Veneziano,
  ``Superstring collisions at planckian energies,''
  Phys.\ Lett.\  B {\bf 197}, 81 (1987).
}
\lref\AmatiUF{
  D.~Amati, M.~Ciafaloni and G.~Veneziano,
  ``Classical and quantum gravity effects from planckian energy superstring 
  collisions,''
  Int.\ J.\ Mod.\ Phys.\  A {\bf 3}, 1615 (1988).
}
\lref\AmatiTN{
  D.~Amati, M.~Ciafaloni and G.~Veneziano,
  ``Can Space-Time Be Probed Below The String Size?,''
  Phys.\ Lett.\  B {\bf 216}, 41 (1989).
}
\lref\GrossAR{
  D.~J.~Gross and P.~F.~Mende,
  ``String theory beyond the Planck scale,''
  Nucl.\ Phys.\  B {\bf 303}, 407 (1988).
}
\lref\BanksGD{
  T.~Banks and W.~Fischler,
  ``A model for high energy scattering in quantum gravity,''
  arXiv:hep-th/9906038.
}
\lref\EardleyRE{
  D.~M.~Eardley and S.~B.~Giddings,
  ``Classical black hole production in high-energy collisions,''
  Phys.\ Rev.\  D {\bf 66}, 044011 (2002)
  [arXiv:gr-qc/0201034].
}
\lref\SGII{
  S.~B.~Giddings,
  ``Locality in quantum gravity and string theory,''
  Phys.\ Rev.\  D {\bf 74}, 106006 (2006)
  [arXiv:hep-th/0604072].
  }
  \lref\SGMWIP{
 S.~B.~Giddings, D.~J.~Gross and A.~Maharana,
  ``Gravitational effects in ultrahigh-energy string scattering,''
  arXiv:0705.1816 [hep-th].
} 
  \lref\GiddingsIE{
  S.~B.~Giddings,
  ``Quantization in black hole backgrounds,''
  arXiv:hep-th/0703116.
}
\lref\SGI{
  S.~B.~Giddings,
``Black hole information, unitarity, and nonlocality,''
  Phys.\ Rev.\  D {\bf 74}, 106005 (2006)
  [arXiv:hep-th/0605196].
}
\lref\SGinfo{S.~B.~Giddings,
  ``Quantum mechanics of black holes,''
  arXiv:hep-th/9412138\semi
  ``The black hole information paradox,''
  arXiv:hep-th/9508151.
}
\lref\Astroinfo{
  A.~Strominger,
  ``Les Houches lectures on black holes,''
  arXiv:hep-th/9501071.
}
\lref\Pageinfo{
  D.~N.~Page,
  ``Black hole information,''
  arXiv:hep-th/9305040.
}
%
%
\lref\BanksCG{
  T.~Banks,
  ``Some Thoughts on the Quantum Theory of de Sitter Space,''
  arXiv:astro-ph/0305037.
}
\lref\BanksWR{
  T.~Banks, W.~Fischler and S.~Paban,
  ``Recurrent nightmares?: Measurement theory in de Sitter space,''
  JHEP {\bf 0212}, 062 (2002)
  [arXiv:hep-th/0210160].
}
\lref\Bankslittle{
T.~Banks,
``Cosmological breaking of supersymmetry or little Lambda goes back to  the future. II,''
arXiv:hep-th/0007146.
}
\lref\GiLitwo{
  S.~B.~Giddings and M.~Lippert,
  ``The information paradox and the locality bound,''
  Phys.\ Rev.\ D {\bf 69}, 124019 (2004)
  [arXiv:hep-th/0402073].
}
\lref\GiLione{
  S.~B.~Giddings and M.~Lippert,
  ``Precursors, black holes, and a locality bound,''
  Phys.\ Rev.\ D {\bf 65}, 024006 (2002)
  [arXiv:hep-th/0103231].
}
\lref\PageSC{
  D.~N.~Page,
  ``Is Black Hole Evaporation Predictable?,''
  Phys.\ Rev.\ Lett.\  {\bf 44}, 301 (1980).
}
\lref\Boltz{L. Boltzmann, ``On certain questions of the theory of gases," Nature {\bf 51}, 413 (1895).}
\lref\Rees{M.~J.~Rees, {\sl Before the beginning: our universe and others} (Simon and Schuster, New York, 1997), p 221.}
\lref\AlSo{
  A.~Albrecht and L.~Sorbo,
  ``Can the universe afford inflation?,''
  Phys.\ Rev.\  D {\bf 70}, 063528 (2004)
  [arXiv:hep-th/0405270].
}
\lref\PageBB{
  D.~N.~Page,
 ``Is our universe likely to decay within 20 billion years?,''
  arXiv:hep-th/0610079.
}
\lref\BoussoXC{
  R.~Bousso and B.~Freivogel,
  ``A paradox in the global description of the multiverse,''
  arXiv:hep-th/0610132.
}
\lref\BanksTA{
  T.~Banks and W.~Fischler,
  ``Holographic cosmology 3.0,''
  Phys.\ Scripta {\bf T117}, 56 (2005)
  [arXiv:hep-th/0310288].
}
\lref\BanksRX{
  T.~Banks, B.~Fiol and A.~Morisse,
 ``Towards a quantum theory of de Sitter space,''
  JHEP {\bf 0612}, 004 (2006)
  [arXiv:hep-th/0609062].
}
\lref\ArkaniHamedKY{
  N.~Arkani-Hamed, S.~Dubovsky, A.~Nicolis, E.~Trincherini and G.~Villadoro,
  ``A Measure of de Sitter Entropy and Eternal Inflation,''
  arXiv:0704.1814 [hep-th].
}
\lref\UQM{S.B. Giddings, ``Universal quantum mechanics," {\sl in preparation}.}
\Title{\vbox{\baselineskip12pt
}}
{\vbox{\centerline{Black holes, information, and locality}
}}
\centerline{{\ticp Steven B. Giddings}\footnote{$^\star$}{Email address: giddings@physics.ucsb.edu} 
}
\centerline{ \sl Department of Physics}
\centerline{\sl University of California}
\centerline{\sl Santa Barbara, CA 93106-9530}
\bigskip
\centerline{\bf Abstract}

Thirty years of a deepening information paradox suggest the need to revise our basic physical framework.  Multiple indicators point toward reassessment of  the principle of locality: lack of a precise definition in quantum gravity, behavior of high-energy scattering, hints from strings and AdS/CFT, conundrums of quantum cosmology, and finally lack of good alternative resolutions of the paradox.  A plausible conjecture states that the non-perturbative dynamics of gravity is unitary but nonlocal.  String theory may directly address these issues but so far  important aspects remain elusive.  If this viewpoint is correct, critical questions are to understand the ``correspondence" limit where nonlocal physics reduces to local quantum field theory, and beyond, to unveil principles of an underlying nonlocal theory.

\Date{}

Physics periodically confronts circumstances requiring major conceptual redirection. Thirty years of a deepening black hole information paradox strongly suggests it 
exemplifies such a confrontation.  While the next conceptual step is not fully apparent, there are significant indications of the need to  substantially revise  local quantum field theory.

The essential paradox is the statement that any proposal for the fate of information fallen into a black hole contradicts a cherished principle of physics.  Hawking's proposal that black holes evaporate completely, destroying their information\Hawkunc\ not only eliminates unitary quantum evolution, but was also argued\refs{\EHNS,\BPS} to  imply massive violations of energy conservation, raising ambient vacuum temperatures to the Planck scale $M_p$.   If black holes instead cease evaporating near $M_p$, leaving behind remnants, these should have infinite internal states encoding information from arbitrarily large initial holes.  Stability is sacrificed: quantum arguments indicate remnants would be infinitely produced in generic physical processes.  The third alternative -- information is radiated long before this stage -- apparently conflicts with macroscopic locality, since information must propagate outside the light cone.\foot{For more complete reviews, see \refs{\SGinfo\Astroinfo-\Pageinfo}.}

The simplicity and depth of the paradox suggest comparison with other crises in physics and the birth of quantum physics seems  apt.   There physicists needed to abandon certain concepts of classical physics, particularly the notion of deterministic evolution on phase space.  One is led to ask for current analogs.

A sense has grown that { \it macroscopic locality} should not be a fundamental property of physics; early proposals in this direction include\refs{\BHMR\tHooholo-\Sussholo}.  There are several reasons to question locality: lack of a precise definition in quantum gravity -- connected with the apparent absence of local observables; indications from high-energy gravitational scattering; hints from string theory, particularly AdS/CFT; conundrums of quantum cosmology; and finally the paradox itself and the appearance that other outs are even more radical.

Locality is a basic axiom of quantum field theory(QFT), stated as commutativity of gauge-invariant local observables outside the light cone.  In gravity, the problem begins in defining such observables.  Any local operator transforms non-trivially under diffeomorphisms
\eqn\diffobs{\delta_\xi {\cal O}(x) = \xi^\mu \partial_\mu {\cal O}(x)\neq 0}
so is not gauge-invariant.  

The idea of {\it relational}  observables traces through many works\refs{\EinsDov\BryceI\BryceII\PageUC\BanksCW\HartleGN\RovelliJM\RovelliPH\RovelliPI\RovelliBZ\TsamisYU\SmolinIS\AshtekarWB\MarolfWH\MarolfNZ\AmbjornMK\AmbjornWC\GambiniPE\GambiniYJ\DittrichCB\DittrichKC\GMH\ThiemannWK-\PonsRZ} to Einstein;  localization is defined in relation to features of a background state -- geometry or  other field structure. 
One can apparently formulate diffeomorphism-invariant relational observables that reduce to local QFT observables when quantum gravity is irrelevant;  such constructions were outlined in \GMH.   These {\it proto-local} observables have only limited locality, when the backreaction on the background is not too strong and other conditions are satisfied.  Such limitations, relation to Copenhagen measurements, and the notion of an ``ultimate detector," appear in \GMH.  Toy models for proto-local observables have been precisely formulated in two-dimensional gravity\GaGi, illustrating approximate localization and such limitations.  These investigations suggest that locality is both {\it relative} and {\it approximate.}

High-energy scattering also probes locality.  Very general QFT assumptions, including locality, 
imply bounds on scattering cross sections.  One is the Froissart bound
\eqn\froiss{\sigma_T= ({\rm const.})\log^{D-2} E\ }
on the maximal growth with energy of the total scattering cross section.
In contrast, strong gravitational effects, particularly black hole formation, indicate absorptive cross sections growing as the appropriate power of the Schwarzschild radius defined by the CM energy, 
\eqn\bhcross{\sigma_T\propto E^{D-2\over D-3}}
in $D$-dimensions.   This rapid growth in size suggests nonlocal behavior\refs{\SGMWIP}.

In short, there is no apparent fundamental quantum-gravitational definition of locality analogous to that in QFT, and probing locality via scattering exhibits  violation of quite generic bounds on local theories, corresponding to long-distance effects.

String theory  strongly suggests a non-local perspective, particularly through the proposed AdS/CFT correspondence\Mald.   If this  holds at the needed level of detail, it is intrinsically nonlocal, as it describes  $D$-dimensional string theory in AdS in terms of a $(D-1)$-dimensional  gauge theory residing on its boundary.   A closely related generalization is the conjectured {\it holographic principle}\refs{\tHooholo,\Sussholo,\BoussoJU}.  One may divide this into two versions.  The strong holographic principle asserts that physics in a region of spacetime is equivalently described by a theory on its boundary.  A weaker version states that the number of degrees of freedom in a region is proportional to an appropriately-defined boundary area.  Both conflict with local QFT.

As a proposed concrete realization of a non-local theory, string theory should be carefully probed.  So far it has resisted yielding clear explanations.  AdS/CFT and related  constructions represent the currently most promising non-perturbative framework.  However, at present AdS/CFT does not clearly reveal how one {\it approximately} recovers local physics at scales between the string length $M_s^{-1}$ and the AdS radius.  AdS/CFT encodes a type of S-matrix\GiddingsQU, but so far its local detail is elusive. Assuming AdS/CFT accurately describes formation and evaporation of a small black hole,  unitary evolution is expected. But we haven't yet understood approximately local (proto-local)
string theory observables  (though suggestions appear in \GaGi).  Thus string theory currently evades a central aspect of the paradox -- reconciling observations of observers falling into a black hole with those of asymptotic observers.  

One can also investigate high-energy string scattering to probe locality.  Consider two-string collisions at ultrahigh  energy with gradually decreasing impact parameter.  Any stringy nonlocality -- {\it e.g.} from string extendedness --  should reveal itself in significant corrections to QFT predictions.  This might be expected when momentum transfers reach $M_s$, or when the strings become appreciably stretched through mutual gravitation.  However, closer investigation\refs{\SGII,\SGMWIP}, following \refs{\AmatiWQ\AmatiUF\AmatiTN\GrossAR\BanksGD-\EardleyRE}, suggests that the important corrections to local QFT occur when gravity gets strong, as the impact parameter approaches the Schwarzschild radius.  These statements apparently indicate that a central question is breakdown of gravitational perturbation theory,  and it is not clear how strings  would prevent this breakdown nor clearly explain its outcome.  Unfortunately, string theory currently remains remarkably moot on these vital issues.

Whether or not described by string theory, it is very important to characterize possible nonlocality.  More complete physical frameworks typically incorporate predecessors through correspondence.  Specifically, any nonlocal framework should reduce to local and causal QFT in an appropriate limit. Understanding this limit is important.  The uncertainty principle parametrizes the region where classical phase space yields to Hilbert space.  In this spirit, one might ask when the Fock space of local QFT yields to something different.  One approach is the  ``locality bound" proposal\refs{\SGIII,\SGII,\GiLione,\GiLitwo}: local field theory ceases to be a good description for states of QFT such that gravitational backreaction becomes non-perturbative, roughly when a black hole is formed.  A related but distinct criterion arises from information bounds of the holographic principle.

One also should understand how nonlocal physics could address black hole information. Two alternatives appear.  Consider the classical model of hydrogen, which becomes singular when the electron spirals into the proton.  Quantum mechanics avoids this breakdown by {\it replacing} classical physics at much larger scales, the atomic radius -- singular dynamics becomes irrelevant.  Likewise local QFT could simply be an inaccurate description of black hole interiors.  Alternatively, local theory could exhibit  explicit breakdown indicating the regime where replacement is needed.  Investigating Hawking evaporation provides hints.  Indeed, arguments have been given\refs{\GiddingsIE,\GiLitwo,\SGI,\SGIII} that the semiclassical approximation, or $1/M_p$ expansion,  breaks down in this context, due to significant couplings between fluctuations, at approximately the required time scale\Page.  There may be other sources of semiclassical breakdown\refs{\PageSC\HorowitzVP-\HorowitzMR}.  If QFT intrinsically fails, a parallel with the history of quantum mechanics is not exact.  In either case, relevance of dynamics on horizon scales  focuses attention away from the singularity -- like with hydrogen.

Either way, the correct viewpoint is proposed to be that the complete description of black holes is beyond local QFT just as  complete description of hydrogen is beyond classical theory.
This fits with a conjectured ``nonlocality principle\refs{\SGI,\SGIII}," stating that the non-perturbative dynamics of gravity is unitary but nonlocal.  If correct, we stand at a point similar to quantum theory pre-1925. We may roughly understand the domain where current theory fails, but not the underlying kinematics and dynamics, analogous to the wavefunction and Schrodinger equation.  The central question is what characterizes a more complete nonlocal theory, especially if string theory doesn't suffice.  Proposals include \refs{\BanksTA,\BanksRX}; one can also  seek clues from other approaches to quantum de Sitter space\SGDM.  A plausible working assumption is that such a theory respects appropriate postulates of quantum mechanics\refs{\UQM}.  A critical question is how such nonlocal dynamics could reduce to usual local, causal physics in appropriate limits.  

Also in the context of eternal inflation and the multitude of apparent vacua of the string landscape
local QFT methods have produced vexing conundrums.  Examples include proliferation of ``Boltzman brain" observers\refs{\Boltz\Rees\AlSo-\PageBB}, and other large numbers.  Plausibly nonlocal modifications regulate some of the problematic behavior, a proposal being investigated from various directions\refs{\BoussoXC,\SGDM}.  Proper description of proto-local observables is likely important for such quantum-cosmological issues.

Sacrificing macroscopic locality seems radical but not in comparison with other principles that would apparently need to be abandoned as alternate resolutions to the information paradox.  Moreover,  there is apparently not a good case for a precise notion of locality in non-perturbative quantum gravity, and there are certainly strong hints of non-locality and the possibility that it could resolve some of our most essential problems.  A central question remains:  what is the underlying ``Nonlocal Mechanics?"

\bigskip\bigskip\centerline{{\bf Acknowledgments}}\nobreak

The author wishes to thank  N. Arkani-Hamed, D. Gross, A. Maharana, D. Marolf, and  J. Hartle for important discussions.
This work  was supported in part by the Department of Energy under Contract DE-FG02-91ER40618,  and by grant 
RFPI-06-18 from the Foundational Questions Institute (fqxi.org).


\listrefs
\end